\documentclass[aps,prl,twocolumn,showpacs,amsmath,amssymb,amsfonts,nofootinbib,long,floatfix]{revtex4}
\usepackage{epsfig,latexsym,bm,longtable}

\newcommand\Mpc{\mbox{Mpc}}

\newcommand{\BS}{\!\!\!\!\!\!\!\!\!\!\!\!\!\!\!}

\begin{document}

\title{TESTING THE ISOTROPY OF THE UNIVERSE WITH TYPE Ia SUPERNOVAE}

\author{L. Campanelli$^{1}$}
\email{leonardo.campanelli@ba.infn.it}

\author{P. Cea$^{1,2}$}
\email{paolo.cea@ba.infn.it}

\author{G. L. Fogli$^{1,2}$}
\email{gianluigi.fogli@ba.infn.it}

\author{A. Marrone$^{1,2}$}
\email{antonio.marrone@ba.infn.it}

\affiliation{$^1$Dipartimento di Fisica, Universit\`{a} di Bari, I-70126 Bari, Italy}
\affiliation{$^2$INFN - Sezione di Bari, I-70126 Bari, Italy}

\date{March, 2011}

%************************************   Abstract   *******************************************%

\vspace*{0.3cm}

\begin{abstract}
\begin{center}
{\bf Abstract}
\end{center}
We analyze the magnitude-redshift data of type Ia supernovae included in the Union and Union2
compilations in the framework of an anisotropic Bianchi type I cosmological model and in the
presence of a dark energy fluid with anisotropic equation of state. We find that the amount
of deviation from isotropy of the equation of state of dark energy, the skewness $\delta$,
and the present level of anisotropy of the large-scale geometry of the Universe, the actual
shear $\Sigma_0$, are constrained in the ranges $-0.16 \lesssim \delta \lesssim 0.12$ and
$-0.012 \lesssim \Sigma_0 \lesssim 0.012$ ($1\sigma$ C.L.) by Union2 data.
Supernova data are then compatible with a standard isotropic universe ($\delta = \Sigma_0 = 0$),
but a large level of anisotropy, both in the geometry of the Universe and in the equation of
state of dark energy, is allowed.
\end{abstract}

%*********************************************************************************************%

%98.80.Es -> Observational cosmology
%98.80.Jk -> Mathematical and relativistic aspects of cosmology
\maketitle

%***********************************   Introduction   ****************************************%

\section{\normalsize{I. Introduction}}
\renewcommand{\thesection}{\arabic{section}}

Probing the large-scale geometry of the Universe at cosmological scales is, undoubtedly,
one of the most outstanding issues in modern cosmology. The standard assumptions of
homogeneity and isotropy (namely, the Cosmological Principle~\cite{Weinberg}) can now be tested
via new and very accurate data coming from the study of cosmic microwave background (CMB)
radiation, especially from the Wilkinson Microwave Anisotropy Probe (WMAP)~\cite{WMAP7},
and data from type Ia supernovae, such as those collected in the so-called Union~\cite{Kowalski}
and Union2~\cite{Union2} compilations.
Indeed, concerning the tests of isotropy,
an anisotropic model of Universe, known as ``ellipsoidal universe''~\cite{prl,prd,prd2,ADE,parallax,Cea1,Cea2},
may be even favored by an observed anomalous feature of the CMB power spectrum
--the lack of power on large angular scales-- while being consistent with other cosmological data.

It is not excluded that such an anisotropic model of the Universe could even account for
three other large-scale ``anomalies'' of the isotropic standard cosmological model
(for a brief but pointed discussion see Ref.~\cite{Antoniou}):
the detection of large-scale velocity flows significatively larger than those predicted in standard
cosmology~\cite{A1}, a statistically significant alignment and planarity of the CMB
quadrupole and octupole modes~\cite{A2}, and
the observation of large-scale alignment in quasar polarization vectors~\cite{A3}.
It should be stressed, however, that the above large-angle anomalies
are still subject to an intense debate, since they could be indeed related
to some common systematic.

In this paper, we present an analysis of the large-scale isotropy assumption by means of
magnitude-redshift data of type Ia supernovae (SNe). In particular, we use data from both Union
and Union2 compilations,
consisting of 307 and 557 type Ia SNe respectively, to set constraints on the parameters of an anisotropic model of
the Universe
(for earlier work on the possibility to test the Cosmological Principle with SN data,
see Ref.~\cite{Kolatt,Gupta1,Gupta2,Koivisto-Mota}, while for recent works, see Ref.~\cite{Colin,Quartin}
and references therein).

We assume an anisotropic Bianchi type I cosmological model~\cite{Bianchi},
characterized by a cosmic shear $\Sigma$
in the presence of a dark energy fluid with anisotropic equation of state,
characterized by a skewness $\delta$.
This fluid, firstly studied by Barrow in Ref.~\cite{Barrow},
could be produced by the dynamics of a cosmic vector field,
as shown by Koivisto and Mota in Ref.~\cite{Koivisto-Mota}.
Other effects could give rise to an ellipsoidal universe, such
as a large-scale cosmic magnetic field~\cite{prl,prd,prd2},
or a dark energy fluid having a nonvanishing velocity with respect to the CMB frame~\cite{Jimenez}.
For an incomplete list of such mechanisms of universe anisotropization see, e.g.,
Ref.~\cite{Antoniou} and references therein.

Testing the dark energy anisotropic model with type Ia SNe implies that we can only constrain the anisotropy
parameters ($\Sigma$ and $\delta$) at relatively recent times (i.e., at redshift $z \lesssim 1.6$),
their earlier evolution being largely unconstrained.

We find no evidence in favor of anisotropies of either geometric origin ($\Sigma \neq 0$)
or dark-energy origin ($\delta \neq 0$). However, we can put significant upper and lower bounds
on the deviations of $\Sigma$ and $\delta$ from zero.

The paper is organized as follows. In the next Section we set up the formalism
of a cosmological model with anisotropic fluid while, in Section III, we derive the
magnitude-redshift relation for such a universe.
In Section IV, we use magnitude-redshift data of SNe from the Union and Union2 compilations
to constrain all the free parameters of the model, including $\Sigma$ and $\delta$,
so as to test the isotropy of the observable universe at $z \lesssim 1.6$.
In Section V, we draw our conclusions.

%*********************************************************************************************%

\section{\normalsize{II. Anisotropic Cosmological Model: Ellipsoidal Universe}}
\renewcommand{\thesection}{\arabic{section}}

In order to test possible anisotropies of the Universe we need
to make assumptions beyond
the standard cosmological model
(the Lambda cold dark matter ($\Lambda$CDM) concordance model~\cite{Weinberg}),
which is isotropic.

For the sake of simplicity, in this paper we
consider an anisotropic Bianchi I type cosmological model
with the highest (planar) symmetry in the spatial sections of the spacetime geometry.
Named as ``ellipsoidal universe'' in~\cite{prl,prd,prd2,ADE,parallax,Cea1,Cea2}, it has the
attractive feature of accounting for the
observed lack of power of the cosmic microwave background anisotropy at large scales.

In this model, the most general plane-symmetric line element is~\cite{Taub}:
\begin{equation}
\label{metric}
ds^2 = dt^2 - a^2(t) (dx^2 + dy^2) - b^2(t) \, dz^2\ ,
\end{equation}
where $a$ and $b$ are the scale factors (or expansion parameters),
which we normalize as $a(t_0) = b(t_0) = 1$ at the present time $t_0$.
\footnote{In the standard cosmological model, $a=b$ at all times.}
The mean Hubble parameter $H$ is defined as
\begin{equation}
H \equiv \frac{\dot{A}}{A} \ ,
\end{equation}
$A \equiv (a^2b)^{1/3}$ being the mean expansion parameter, while the
Hubble parameter in the symmetry plane is
\begin{equation}
H_a \equiv \frac{\dot{a}}{a} \ .
\end{equation}
Here and in the following a dot denotes
differentiation with respect to the cosmic time.
The cosmic shear $\Sigma$ is defined as:
\begin{equation}
\label{ShearHubble} \Sigma \equiv \frac{H_a - H}{H} \ .
\end{equation}
In an ellipsoidal universe, the most general energy-momentum tensor
compatible with the metric in Eq.~(\ref{metric}) is of the form
\begin{equation}
\label{tensor}
T^{\mu}_{\,\,\, \nu} = \mbox{diag} \, (\rho,-p_{\|},-p_{\|},-p_{\bot})\ ,
\end{equation}
where $p_{\|}$ and $p_{\bot}$ are ``longitudinal'' and ``normal'' pressures.
In Friedmann universes, $p_{\|} = p_{\bot}$, reflecting the isotropy
of the metric. Conversely, anisotropic universes can support fluids with $p_{\|} \neq p_{\bot}$,
which we will refer to as anisotropic fluids.

Given the above energy-momentum tensor, Einstein's equations read
\begin{eqnarray}
\label{E1} && \BS \, 3(1-\Sigma^2) H^2 = 8\pi G \rho\ , \\
\label{E2} && \BS \, 3(1-\Sigma + \Sigma^2) H^2 + \frac{d}{dt} [(2-\Sigma) H] =
           -8\pi G p_{\|}\ , \\
\label{E3} && \, \BS 3(1+\Sigma)^2 H^2 + 2  \frac{d}{dt}[(1+\Sigma) H] =
           -8\pi G p_{\bot}\ ,
\end{eqnarray}
where $G$ is the Newton constant.

By linearly combining Einstein's equations we get
\begin{equation}
\label{EqShear} \frac{d}{dt} (H\Sigma) + 3H^2\Sigma =
\frac{8\pi G}{3} (p_{\|} - p_{\bot})\ ,
\end{equation}
which shows  that asymmetric pressures  can act as a source of the shear, and
\begin{equation}
\label{EqEnergy} \dot{\rho} + 3H \left( \rho + \frac{2p_{\|} + p_{\bot}}{3} \right)
+ 2H (p_{\|} - p_{\bot}) \Sigma = 0\ ,
\end{equation}
which represents the time component of the energy-momentum conservation law,
$T^{\mu}_{\,\,\, \nu \, ;\mu} = 0$.

We assume that the anisotropic fluid defined in Eq.~(\ref{tensor}) is indeed made up by
an isotropic, pressureless dark matter (DM) component and an anisotropic, dark energy (DE)
component, with equations of state
\begin{equation}
p_{\|} \equiv w_{\|} \, \rho_{\rm DE} \ , \;\;\; p_{\bot} = w_{\bot} \, \rho_{\rm DE} \ ,
\end{equation}
where $\rho_{\rm DE}$ is the dark energy density.
For the sake of simplicity, we assume constant $w_{\|}$ and $w_{\bot}$ coefficients.

In terms of $w_{\|}$ and $w_{\bot}$, we define a mean coefficient $w$ as
\begin{equation}
w \equiv \frac{2w_{\|} + w_{\bot}}{3} \ ,
\end{equation}
and a skewness $\delta$ as
\begin{equation}
\delta \equiv w_{\|} - w_{\bot}\ .
\end{equation}
%
%Let us observe that $w_{\|} = w + \delta/3$ and $w_{\bot} = w - 2\delta/3$.
%
We also assume noninteracting DM and DE fluids, so that the isotropic and anisotropic
components are separately conserved. Writing
\begin{equation}
\rho = \rho_m + \rho_{\rm DE}\ ,
\end{equation}
where $\rho_m$ is the dark matter density,
Eq.~(\ref{EqEnergy}) gives
\begin{eqnarray}
\label{EqDM} && \rho_m = \rho_m^{(0)} A^{-3}\ , \\
\label{EqDE} && \dot{\rho}_{\rm DE} + \left[ 3(1+w) + 2\delta \Sigma \right] H \rho_{\rm DE} = 0\ ,
\end{eqnarray}
where the superscript ``0'' denotes quantities evaluated at the present time $t_0$.

We finally recast Einstein's equations in terms of dimensionless quantities as
\begin{eqnarray}
\label{Eq1} && \bar{H} A \frac{d(\bar{H} \Sigma)}{dA} + 3\bar{H}^2  \Sigma = \delta \Omega_{\rm DE} \bar{\rho}_{\rm DE}\ , \\
\label{Eq2} && A \frac{d\bar{\rho}_{\rm DE}}{dA}  + \left[ 3(1+w) + 2\delta \Sigma \right] \bar{\rho}_{\rm DE} = 0\ ,
\end{eqnarray}
with
\begin{equation}
\label{EqH} \bar{H} = \sqrt{\frac{\Omega_m A^{-3} + \Omega_{\rm DE} \bar{\rho}_{\rm DE}}{1-\Sigma^2}}\ .
\end{equation}
In Eqs.~(\ref{Eq1})--(\ref{EqH}), each barred quantity is normalized to its actual at $t_0$, and the usual
energy density parameters are introduced,
\begin{equation}
\Omega_m \equiv \frac{\rho_m^{(0)}}{\rho_{\rm cr}^{(0)}} \ , \;\;\;
\Omega_{\rm DE} \equiv \frac{\rho_{\rm DE}^{(0)}}{\rho_{\rm cr}^{(0)}} \ ,
\end{equation}
the critical density being at $t_0$
\begin{equation}
\rho_{\rm cr}^{(0)} \equiv \frac{3H_0^2}{8\pi G} \ .
\end{equation}
It is worth noting that, given $\Omega_m$ and $\Sigma_0$, $\Omega_{\rm DE}$ is derived
from Eq.~(\ref{EqH}) as
\begin{equation}
\label{OmegaDE} \Omega_{\rm DE} = 1 - \Omega_m - \Sigma_0^2\ .
\end{equation}
For given parameters $\{\Omega_m, w, \delta, \Sigma_0 \}$,
Eqs.~(\ref{Eq1})--(\ref{Eq2}) are solved numerically. The resulting solution,
once the value of $H_0$ is fixed, describes the evolution of the ellipsoidal universe.

%*********************************************************************************************%

\section{\normalsize{III. Distance Modulus in Ellipsoidal Universe}}
\renewcommand{\thesection}{\arabic{section}}

Within the anisotropic metric~(\ref{metric}),
the luminosity distance $d_L$ of a source at redshift $z$, seen along the direction
$\hat{p}$ ($|\hat{p}| = 1$), is given by~\cite{Koivisto-Mota,Koivisto-Mota2}
\begin{equation}
\label{dL} d_L(z,\hat{p}) = (1+z) \int_{t(z)}^{t_0} \frac{dt}{\left( \, \sum_i a_i^2 \hat{p}_i^2 \right)^{1/2}}\ ,
\end{equation}
where $a_1 \equiv a_2 \equiv a$ and $a_3 \equiv b$ and~\cite{Trodden,Koivisto-Mota,Koivisto-Mota2}
\begin{equation}
\label{z} 1 + z = \left( \sum_i \frac{\hat{p}_i^2}{a_i^2} \right)^{\!\!1/2}\ .
\end{equation}
The above result applies to the case
where the axis of symmetry is identified with the $z$-axis.
To account for a symmetry axis directed
along a generic direction $(b,l) = (b_A,l_A)$ in
galactic coordinates ($b$ and $l$ being
the galactic latitude and longitude, respectively),
we perform rotations of the coordinate system along $z$ and $x$,
\begin{equation}
\label{Rotation} \mathcal{R} \equiv \mathcal{R}_{x}(\pi/2-b_A) \,
\mathcal{R}_{z}(\pi/2+l_A)
\end{equation}
where the arguments are the rotation angles.
In the galactic coordinate system, the
direction cosines of the symmetry axis are
\begin{equation}
\label{nA} \hat{n}_A = (\cos\!b_A \cos l_A, \cos\!b_A \sin l_A, \sin\!b_A)\ ,
\end{equation}
while the components of the generic direction $\hat{p}$ are defined by
direction cosines $\hat{n} = \mathcal{R}^{-1} \hat{p}$, namely,
\begin{equation}
\label{n} \hat{n} = (\cos\!b \cos l, \cos\!b \sin l, \sin\!b)\ .
\end{equation}
The angle $\theta$ between $\hat{n}$ and $\hat{n}_A$ is defined by
\begin{equation}
\label{cos} \cos \theta \equiv \hat{n} \cdot \hat{n}_A\ .
\end{equation}
It is useful to introduce the ``eccentricity'' $e$ as
\begin{equation}
\label{eccentricity} e^2 \equiv 1 - \frac{b^2}{a^2}\ ,
\end{equation}
which is connected to the shear via Eq.~(\ref{ShearHubble}) as:
\begin{equation}
\label{eccentricityA} e^2 = 1 - \exp \! \left[ \, 6 \! \int_A^1 \frac{dx}{x} \, \Sigma(x) \right] \! .
\end{equation}

In galactic coordinates system, the redshift and distance modulus read then
\begin{equation}
\label{znew} 1 + z = \frac1A \frac{\left(1 - e^2 \sin^2\!\theta \right)^{1/2}}
                                  {\left( 1-e^2 \right)^{1/3}}\ ,
\end{equation}
and
\begin{equation}
\label{dLnew} d_L(z,\theta) = \frac{1+z}{H_0}
\int_{A(z)}^{1} \frac{dA}{A^2 \bar{H}} \frac{\left( 1-e^2 \right)^{1/6}}
{\left( 1 - e^2 \cos^2\!\theta \right )^{1/2}}\ ,
\end{equation}
where $\bar{H}$ and $e$ are taken as function of $A$, and $A(z)$ is the solution of Eq.~(\ref{znew}).

Finally, we introduce the usual distance modulus $\mu$, as
\begin{equation}
\label{mu} \mu = 5 \, \log_{10} \!\! \left( \frac{d_L}{1 \, \Mpc} \right) + 25\ .
\end{equation}
It depends on two different sets of parameters, $\mu = \mu(i,j)$,
where $i = \{ z, b, l \}$ characterizes the position of the supernova, while
$j = \{ H_0, \Omega_m, \Sigma_0, w, \delta, b_A, l_A \}$ characterizes the cosmological model.

In the next section, we use the magnitude-redshift SN data to
constrain the free parameters of the the ellipsoidal universe model with anisotropic
dark energy.

%*********************************************************************************************%

\section{\normalsize{IV. Type Ia Supernovae: Constraining the level of Cosmic Anisotropy}}
\renewcommand{\thesection}{\arabic{section}}

%************************************   Figure 1   *******************************************%

\begin{figure*}[t]
\begin{center}
\hspace*{-1.7cm}
\includegraphics[scale=1.1]{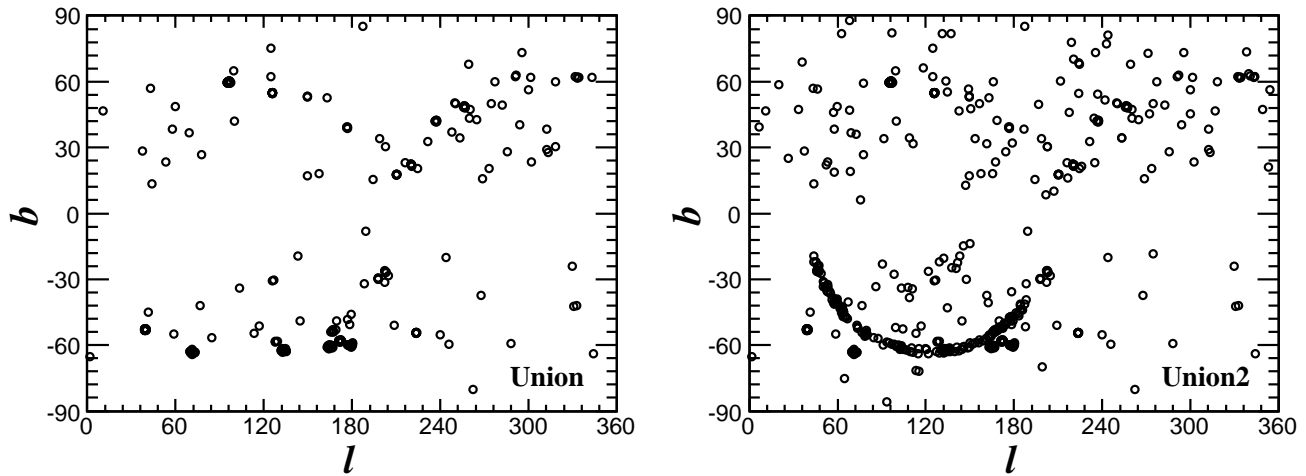}
\vspace*{-1.0cm}
\caption{Angular position of type Ia supernovae in the Union (left panel) and Union2
(right panel) compilations. $b$ and $l$ are the galactic latitude and galactic longitude
of supernovae, respectively (in degrees).}
\end{center}
\end{figure*}

%*********************************************************************************************%

In a given cosmological model, the analysis of the SN lightcurves provides, for each supernova,
the coordinates $(z,\mu)$ which form the so-called Hubble diagram.

In the following, we discuss the Hubble diagram for the Union compilation~\cite{Kowalski}, which consists of 307 supernovae,
and the recently published Union2 compilation~\cite{Union2}, consisting of 557 supernovae.

For our anisotropic cosmological model, the distance modulus depends
(besides the cosmological parameters) on the redshift and angular position of the source.
Hence, in order to fit experimental data, we need for each supernova its angular position.
Figure~1 shows the angular position $(b,l)$ of SNe in the Union and Union2 compilations,
$b$ and $l$ being the galactic latitude and longitude, respectively. Data on angular positions are taken from
Ref.~\cite{Hamuy,Jha,Kowalski,Asiago,Tonry,Barris,harvard,Riess,Blakeslee,Riess2,Astier,Miknaitis,Krisciunas,5SNe}.
The non-uniform distribution of the data (e.g. the absence of SNe around the galactic plane at $b=0$)
reflects observational selection cuts which, however, do not affect our subsequent fit procedure.

\subsection{\normalsize{IV.a Union Compilation}}
\renewcommand{\thesection}{\arabic{section}}

From the observed SN lightcurve, one can infer the distance modulus $\mu_B$
as the difference between the peak bolometric apparent magnitude
$m_B^{\rm max}$ and the absolute magnitude $M$ of supernova. The derived
distance modulus, as well known, is affected by the
so-called stretch and color corrections, $s$ and $c$ respectively.
The amount of such corrections is adjusted by means of free parameters
$\alpha$ and $\beta$ as~\cite{Kowalski}
\begin{equation}
\label{muB} \mu_B = m_B^{\rm max} - M + \alpha(s-1) - \beta c\ ,
\end{equation}
with $M$, $\alpha$ and $\beta$ to be determined from fits to data.
The experimental values $m_B^{\rm max}$, $s$, and $c$ for each supernova
are taken from Ref.~\cite{Kowalski}.

In order to constrain the free parameters of our anisotropic cosmological model,
we compare the experimental distance modulus
$\mu_B(M,\alpha,\beta)$ with the theoretical expectation
$\mu(H_0, \Omega_m, \Sigma_0, w, \delta, b_A, l_A)$ by means of a least-square fit
\begin{equation}
\label{chi2p} \chi^2 = \sum_{\rm SNe} \left[
\frac{(\mu_B' - \mu)^2}{\sigma_\mu^2 + \sigma_{\rm sys}^2 } \, + \frac{\Delta \!M_1^2}{\sigma_{\Delta \!M_1}^2} +
\frac{\Delta M_2^2}{\sigma_{\Delta \!M_2}^2} \, \right]\ ,
\end{equation}
which we now discuss in detail.

%************************************   Figure 2   *******************************************%
\begin{figure*}[t]
\begin{center}
\includegraphics[clip,width=20cm,bb=50 40 550 400]{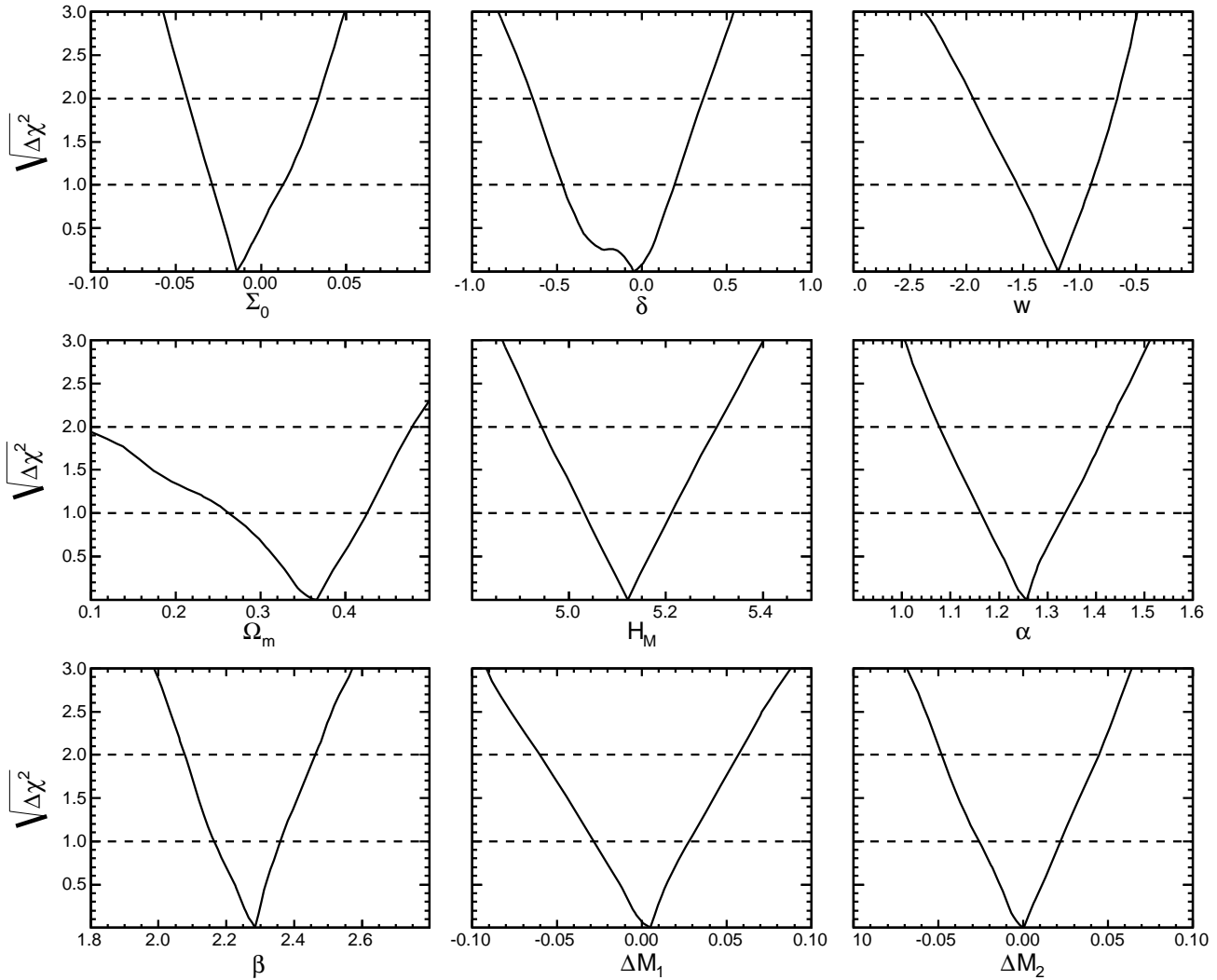}
\caption{$\sqrt{\Delta \chi^2}$ for 9 of the 11 parameters of the
anisotropic cosmological model using the Union set.
The two probability distributions for $l_A$ and $b_A$ are flat (not shown).}
\end{center}
\end{figure*}

%*********************************************************************************************%

{\it Theoretical systematics.}
The uncertainty $\sigma_\mu$ of the theoretical distance modulus $\mu$ contains two
independent contributions,
\begin{equation}
\label{sigmamu} \sigma_\mu^2 = \sigma_{\rm lightcurve}^2 + \sigma_{\rm astro}^2\ .
\end{equation}
The first one, $\sigma_{\rm lightcurve}$, is associated to the fitting procedure of SN lightcurves
and can be expressed as $\sigma_{\rm lightcurve}^2 = \sum_{ij} c_i c_j C_{ij}$~\cite{Kowalski}, where
$c_i = \{ 1,\alpha,-\beta\}$ and $C_{ij}$ is the covariance matrix, whose diagonal parameters are
the uncertainties on peak bolometric apparent magnitude, $\sigma_{m_B}^2$,
stretch, $\sigma_s^2$, and color $\sigma_c^2$.
Since the covariance matrix for each supernova, obtained from the lightcurve fitting procedure
in Ref.~\cite{Kowalski}, is not publicly available, we assume uncorrelated errors on
$m_B$, $s$, and $c$, but a nonzero stretch-color correlation from Ref.~\cite{sc}, $\rho_{sc} = -0.2$.
Accordingly, we get
\begin{equation}
\label{sigmalc} \sigma_{\rm lightcurve}^2 = \sigma_{m_B}^2 + \alpha^2 \sigma_s^2 + \beta^2 \sigma_c^2
+ 2 \rho_{sc} \, \alpha \, \beta \, \sigma_s \sigma_c \ .
\end{equation}
The uncertainty $\sigma_{\rm astro}$ is made up of three
different uncorrelated contributions of astrophysical origin~\cite{Kowalski}:
\begin{equation}
\label{sigmaastro} \sigma_{\rm astro}^2 = \sigma_{\rm v-pec}^2 + \sigma_{\rm lens}^2 + \sigma_{\rm ext}^2\ .
\end{equation}
The first ($\sigma_{\rm v-pec}$) is associated to the peculiar velocities $v_{\rm pec}$ of the host galaxies,
and is estimated by generalizing the analysis of Ref.~\cite{Hui}
(valid in a standard isotropic universe) to the case of an ellipsoidal universe.
We find:
\begin{equation}
\label{sigmavpec} \sigma_{\rm v-pec} = f(z,\theta) \, v_{\rm pec}\ ,
\end{equation}
where
\begin{equation}
\label{f} f(z,\theta) = 1 - \frac{(1+z)^2}{H d_L} \, g(z,\theta)
\end{equation}
with
\begin{equation}
\label{g} g(z,\theta) = \frac{(1-e^2)^{1/2} \, (1-e^2 \sin^2\!\theta)^{1/2} \, (1-e^2 \cos^2\!\theta)^{-1/2}}
{(1-e^2 \sin^2\!\theta) (1+\Sigma) -3\Sigma \cos^2\!\theta}\ .
\end{equation}
Equation~(\ref{sigmavpec}) reduces to the result of Ref.~\cite{Hui} in the limit
of vanishing eccentricity $e$.
As a typical peculiar velocity, we take the value $v_{\rm pec} = 300$km/s.
\\
The second term ($\sigma_{\rm lens}$), is associated to gravitational lensing and can be
parameterized as~\cite{Kowalski}
\begin{equation}
\label{sigmalens} \sigma_{\rm lens} = 0.093 \, z\ .
\end{equation}
Finally, the third term ($\sigma_{\rm ext}$) is associated to Galactic extinction corrections,
and can be estimated as:
\footnote{In Ref.~\cite{Kowalski}, the extinction law of Cardelli et al.~\cite{Cardelli},
together with the $E(B-V)$ values derived from the sky map of Schlegel et al.~\cite{Schlegel},
were used for parameterizing Galactic extinction. Here, we just use that law with the
average Galactic $E(B-V)$.}
\begin{equation}
\label{sigmaextinction} \sigma_{\rm ext} =
\left\{ \begin{array}{ll}
0.013\ , &  z < 0.2\ ,
\\
0\ ,     &  z \geq 0.2\ .
\end{array}
\right.
\end{equation}
{\it Experimental systematics.}
In order to take into account systematic errors on $\mu_B$, we follow the analysis of Ref.~\cite{Kowalski}.
Firstly, we introduce a sample-dependent uncertainty $\sigma_{\rm sys}$ related to
an unknown intrinsic dispersion of the supernova magnitudes, in each of the 13 data subsets
included in the Union compilation. For completeness, these uncertainties are reported in Table~I.
%
%*************************************   Table I   *******************************************%
%
\begin{table}[t]
\begin{center}
\caption{Intrinsic dispersion of the supernova magnitudes, $\sigma_{\rm sys}$,
for each subset of the Union compilation. The second column indicates the number of supernova in each
subset. Data from Ref.~\cite{Kowalski}.} %HST stands for Hubble Space Telescope.
\vspace{0.3cm}
\begin{tabular}{llccc}
\hline \hline
&Subset of Union     &~~SNe &~~$\sigma_{\rm sys}$ &~~Ref. \\
\hline
&Hamuy et al.        &~~17  &~~0.14               &~~\cite{Hamuy}      \\
&Krisciunas et al.   &~~6   &~~0.05               &~~\cite{Krisciunas} \\
&Riess et al.        &~~11  &~~0.16               &~~\cite{Riess1998}  \\
&Jha et al.          &~~15  &~~0.26               &~~\cite{Jha}        \\
&Kowalski et al.     &~~8   &~~0.00               &~~\cite{Kowalski}   \\
&Riess et al. + HST  &~~12  &~~0.28               &~~\cite{Riess1998}  \\
&Perlmutter et al.   &~~29  &~~0.33               &~~\cite{Perlmutter} \\
&Tonry et al.        &~~6   &~~0.06               &~~\cite{Tonry}      \\
&Barris et al.       &~~21  &~~0.23               &~~\cite{Barris}     \\
&Knop et al.         &~~11  &~~0.10               &~~\cite{Knop}       \\
&Riess et al.        &~~27  &~~0.16               &~~\cite{Riess}      \\
&Astier et al.       &~~71  &~~0.12               &~~\cite{Astier}     \\
&Miknaitis et al.    &~~73  &~~0.18               &~~\cite{Miknaitis}  \\
\hline \hline
\end{tabular}
\end{center}
\end{table}

%************************************   Figure 3  *******************************************%
%
\begin{figure*}
\begin{center}
\includegraphics[clip,width=20cm,bb=50 40 570 350]{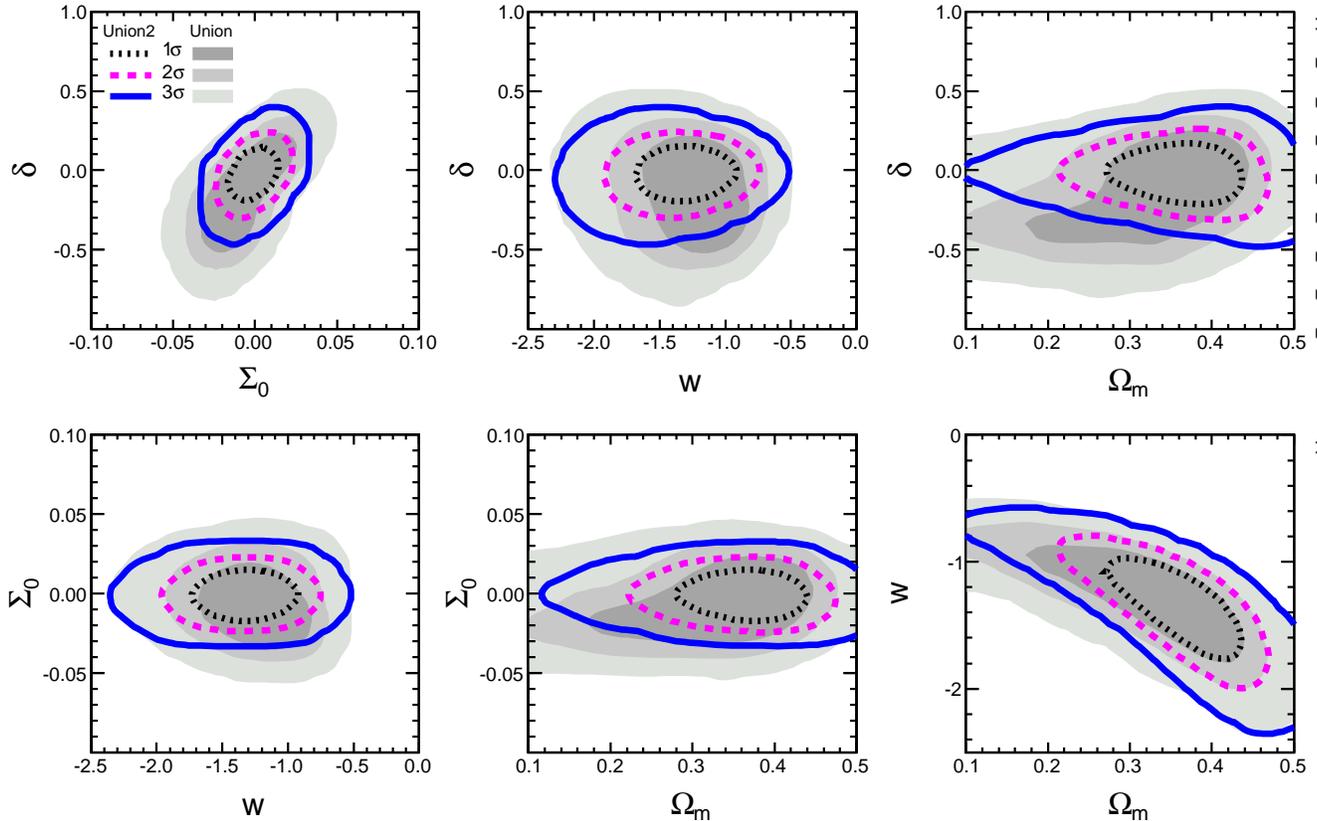}
\caption{$1\sigma$, $2\sigma$, and $3\sigma$ confidence
level contours in the planes $(\Sigma_0,\delta)$, $(w,\delta)$, $(\Omega_m,\delta)$,
$(w,\Sigma_0)$, $(\Omega_m,\Sigma_0)$, and $(\Omega_m,w)$.
Results obtained by using the Union set are shown as gray filled contours, while those coming
from Union2 are represented as empty contours.}
\end{center}
\end{figure*}
%
%*********************************************************************************************%

Secondly, we add two nuisance parameters, $\Delta M_1$ and $\Delta M_2$, shifting $\mu_B$ as
\begin{equation}
\label{muBp} \mu_B' \equiv
\left\{ \begin{array}{ll}
\mu_B + \Delta M_1\ ,              &  z < 0.2\ ,
\\
\mu_B + \Delta M_1 + \Delta M_2\ , &  z \geq 0.2\ .
\end{array}
\right.
\end{equation}
The nuisance parameter $\Delta M_2$ is $z$-dependent, since it accounts for a
possible evolution of supernovae with redshift. The value $z = 0.2$
discriminates, conventionally, between ``low'' and ``high'' redshifts in this context~\cite{Kowalski}.
The uncertainties on the nuisance parameters, $\sigma_{\Delta \!M_1}$ and $\sigma_{\Delta \!M_2}$,
are found by adding in quadrature various uncorrelated uncertainties on
the $\alpha$ and $\beta$ parameters, on contamination, on lightcurve model, on zero point,
on the Malmquist bias, and on galactic extinction. Taking into account the results
of Ref.~\cite{Kowalski} we use the values:
\begin{equation}
\label{sigmaM} \sigma_{\Delta \!M_1} = 0.040\ , \;\;\; \sigma_{\Delta \!M_2} = 0.034\ .
\end{equation}
The previously introduced $\chi^2$ function is then completely defined.

Notice that the two parameters $H_{0}$ and $M$
enter in the $\chi^{2}$ only through a specific combination $H_M$,
\begin{equation}
\label{HM} H_M \equiv \frac{H_0}{\mbox{km}/\mbox{s}/\mbox{Mpc}} \, 10^{-(M+25)/5}\ .
\end{equation}
Therefore, in the analysis we cannot constrain $H$ and $M$ separately, but only $H_{M}$.
We expect $H_M \simeq 5.1$ for small deviations from standard cosmology, where
$H_0 \simeq 71\,\mbox{km}/\mbox{s}/\mbox{Mpc}$
and $M \simeq -19.3$.

Since the $\chi^{2}$ depends on as many as eleven parameters, a brute-force minimization
search on a 11-dimensional grid is not feasible.
We employ then a Markov Chain Monte Carlo approach, which
has become standard in CMB analyses~\cite{WMAP7}. We use a modified version
of the CosmoMC (Cosmological Monte Carlo) code~\cite{Lewis} to produce and analyze the likelihood chains.

{\it Results.}
Figure~2 shows the $\sqrt{\Delta \chi^2}$ distributions for nine parameters
of the model: $w,\Omega_m,H_M,\Sigma_0,\delta,\alpha,\beta,\Delta M_1,\Delta M_2$.
The two remaining parameters $l_A$ and $b_A$, which define the symmetry axis, are
basically unconstrained. This means that there is no evidence for a specific
``anisotropy axis'' from the data (see also discussion in Section IV.c).
However, for any chosen direction, the data can constrain the level of anisotropy,
characterized by the parameters $\Sigma_0$ and $\delta$.

In Fig.~3 (see gray filled contours), we present the $1\sigma$, $2\sigma$, and $3\sigma$ contours in the planes
$(\Sigma_0,\delta)$, $(w,\delta)$, $(\Omega_m,\delta)$, $(w,\Sigma_0)$, $(\Omega_m,\Sigma_0)$,
and $(\Omega_m,w)$.
Here again we follow the convention of Ref.~\cite{PDB}, where the
$1\sigma$, $2\sigma$, and $3\sigma$ joint regions are defined so that the projections
onto each parameter give, respectively,
the $1\sigma$, $2\sigma$, and $3\sigma$ intervals for that particular parameter.
The $\chi^2$ at the best-fit point is 255.2.

An isotropic universe ($\delta = \Sigma_0 = 0$) is consistent with data
and present uncertainties and statistics only allow us to put upper limits
on the level of cosmic anisotropy.

The most correlated variables are $w$ and $\Omega_m$ (bottom-right panel),
and $\delta$ and $\Sigma_0$ (top-left panel).

The anti-correlation between $w$ and $\Omega_m$
is also present in the standard analysis of SN data~\cite{Kowalski} and it
comes from the peculiar dependence of luminosity distance on $w$ and $\Omega_m$.
If one combined SN data with other cosmological information,
such as CMB or Large Scale Structures analyses~\cite{Weinberg},
that degeneracy could be broken so to better constrain $\delta$ and $\Sigma_0$.
However, as it is clear from Fig.~3 [see in particular the $(w,\delta)$ and $(\Omega_m,\delta)$,
and the $(w,\Sigma_0)$ and $(\Omega_m,\Sigma_0)$ panels], the bounds on $\delta$ and $\Sigma_0$
depend only weakly on $w$ and $\Omega_m$.

That $\delta$ and $\Sigma_0$ should be correlated
was to be expected since the deviation from isotropy of the equation of
state of dark energy is the source of the anisotropization of the large-scale geometry of the
Universe. Indeed, in the limit of small anisotropies, i.e. $|\delta| \ll 1$ and $|\Sigma| \ll 1$,
the correlation between $\delta$ and $\Sigma_0$ can be understood as follows.
To the first order in the anisotropies, the solutions of Eqs.~(\ref{Eq1}) and (\ref{Eq2}) are
\begin{equation}
\label{new1}  \Sigma(A) = \frac{\Sigma_0 + (E-E_0) \, \delta}{A^3\bar{H}}
\end{equation}
and
\begin{equation}
\label{new2} \bar{\rho}_{\rm DE} = A^{-3(1+w)}\ ,
\end{equation}
respectively, with the Hubble parameter being
\begin{equation}
\label{new3} \bar{H} = \sqrt{\Omega_m A^{-3} + \Omega_{\rm DE} A^{-3(1+w)}}\ .
\end{equation}
In Eq.~(\ref{new1}) we have introduced the function
\begin{equation}
\label{new4} E(A) = \Omega_{\rm DE} \! \int_0^A \!\! \frac{dx}{x^{1+3w} \bar{H}(x)} \ ,
\end{equation}
and the quantity
\begin{eqnarray}
\label{new5}  E_0 &=& E(1) \nonumber \\
&=& \frac{\sqrt{\Omega_m}}{3w} \! \left( 1 - \frac{1}{\Omega_m} \right)^{\!\frac{1}{2w}}
\!\! B\!\left(\!1 - \frac{1}{\Omega_m};1-\frac{1}{2w},\frac12 \right) \ , \nonumber \\
\end{eqnarray}
where $B(x;a,b)$ is the Incomplete Euler Beta Function and the last equality is valid for $w < 0$.

%************************************   Figure 4  *******************************************%

\begin{figure}[t]
\begin{center}
\vspace*{-0.9cm}
\hspace*{-2.6cm}
\includegraphics[clip,width=0.7\textwidth]{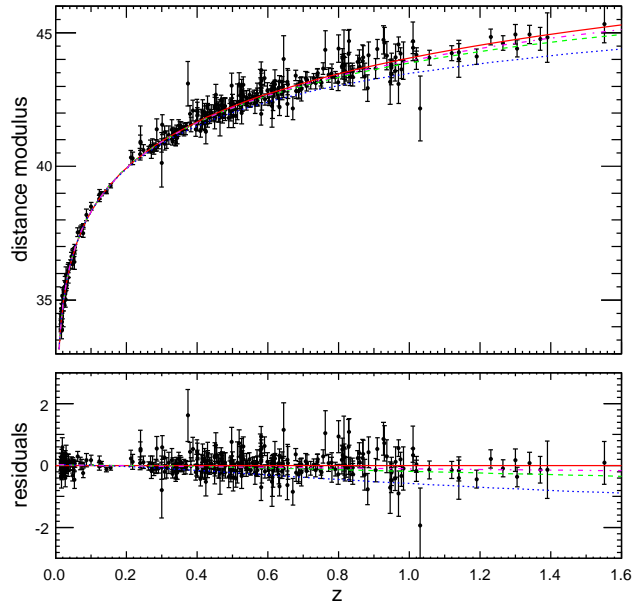}
\vspace*{-1.2cm}
\caption{{\it Upper panel.} Hubble diagram for the 307 supernovae in the Union compilation,
for different cosmological models: best-fit cosmology
$(\Sigma_0,\delta) \simeq (-0.012,-0.013)$ (red continuous line),
$(\Sigma_0,\delta)=(0.2,1)$ (magenta dash-dotted line),
$(\Sigma_0,\delta)=(0.2,0)$ (green dashed line),
$(\Sigma_0,\delta)=(0.2,-1)$ (blue dotted line).
For graphical clarity, we take the SN angular positions and the direction
of the axis of symmetry fixed to the value $(b,l) = (b_A,l_A) = (0,0)$.
The remaining parameters are fixed to their best-fit values.
{\it Lower panel.} Residuals (distance modulus minus distance modulus for the best-fit
cosmology) for the same models in the upper panel.}
\end{center}
\end{figure}

%************************************   Figure 5   *******************************************%
\begin{figure*}[t]
\begin{center}
\hspace*{-1.7cm}
\includegraphics[clip,width=1.23\textwidth]{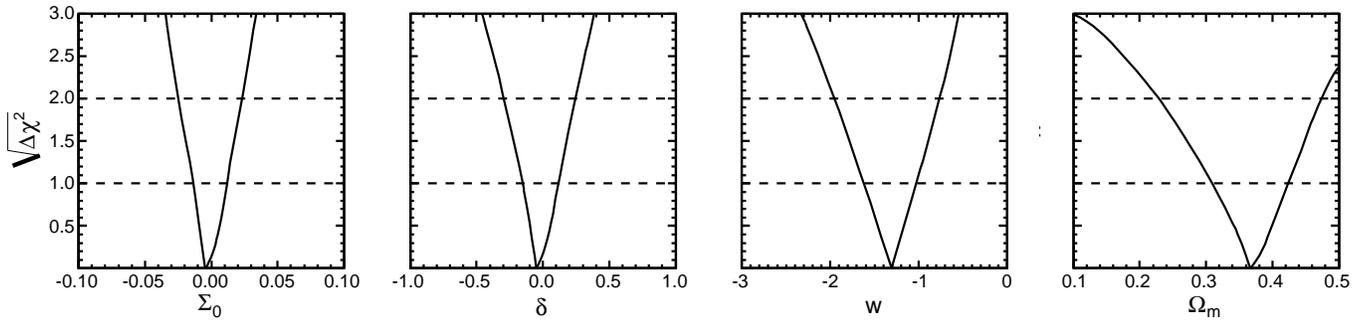}
\vspace*{-0.8cm}
\caption{$\sqrt{\Delta \chi^2}$ for 4 of the 6 parameters of the
anisotropic cosmological model using the Union2 set.
The probability distributions for $l_A$ and $b_A$ are flat (not shown).}
\end{center}
\end{figure*}

%*********************************************************************************************%

At early times ($A \ll 1$), the anisotropic source associated
to dark energy is inefficient to generate an anisotropization of the universe since
the dark energy density is completely negligible with respect to that of the
isotropic dark matter component.
This condition of isotropization at early times, namely
\begin{equation}
\label{new6} \lim_{A \rightarrow 0} \Sigma(A) = 0\ ,
\end{equation}
when applied to Eq.~(\ref{new1}) gives
\begin{equation}
\label{new7}  \delta = \frac{\Sigma_0}{E_0}\ .
\end{equation}
Observing that $E_0$ is a positive-defined quantity which depends only on $w$ and $\Omega_m$,
we find that $\delta$ and $\Sigma_0$ are, indeed, positively correlated.
$E_0$ is of order of $10^{-1}$ for a fiducial cosmology $(\Omega_m,w) = (0.3,-1)$. So,
we expect $\delta \sim 10 \Sigma_0$, and this is confirmed by our data analysis
(see the top-left panel in Fig.~3 and, below, Table~II).

Finally, we observe that the $\alpha$ and $\beta$ best-fit values are
in good agreement with those found in Ref.~\cite{Kowalski} for an isotropic universe,
and that the probability distribution for the nuisance parameters $\Delta M_1$ and $\Delta M_2$
are centered around zero, indicating that no systematic shift of the distance modulus
is needed to fit data. Also, as expected, we find $H_M \simeq 5.1$.

For completeness, we present in Fig.~4 (upper panel) the Hubble diagram
for different realizations of our cosmological model.
For graphical clarity, we take the SN angular positions and the direction
of the axis of symmetry fixed to the value $(b,l) = (b_A,l_A) = (0,0)$.
In the lower panel we show the corresponding distance
modulus fit residual (distance modulus minus the best-fit distance modulus)
as a function of the redshift.
The red continuous line corresponds to our best-fit cosmology, practically
indistinguishable from the one corresponding to the standard isotropic universe (not shown).
%the remaining parameters being fixed to their best-fit values.
%
For a relatively large value $\Sigma_0 = 0.2$
(that we choose for illustrative purposes) and for $\delta=0$,
the distance modulus curve moves downwards (green dashed line).
By taking $\delta = 1$, the distance modulus curve moves upwards
(magenta dash-dotted line), while the opposite happens taking
$\delta = -1$ (blue dotted line).
A similar behavior is found for negative value of the shear, according to the positive
correlation found between $\Sigma_0$ and $\delta$.

\subsection{\normalsize{IV.b Union2 Compilation}}
\renewcommand{\thesection}{\arabic{section}}

The analysis of SN lightcurves in the Union2 compilation
gives the distance modulus $\mu_B$ as~\cite{Union2}
\begin{equation}
\label{new8} \mu_B = m_B^{\rm max} - M + \alpha' x_1 - \beta' c\ ,
\end{equation}
where $m_B^{\rm max}$ and $c$ are, as before, the peak bolometric apparent magnitude and the color correction,
while $x_1$ derives from the fit to the SN lightcurves with the so-called SALT2 (Spectral Adaptive Light curve Template) 
fitter~\cite{SALT2}. $\alpha'$ and $\beta'$ are, instead, nuisance parameters to be determined, simultaneously with
the cosmological parameters and the absolute magnitude $M$, from fits to data.

{\it $\chi^2$ statistics.}
As in the case of Union compilation, the covariance matrices resulting from the lightcurve fit
are not publicly available, so we do not have any information on the correlation between the errors
on $m_B^{\rm max}$, $x_1$, and $c$. Consequently, we follow the analysis
of Ref.~\cite{Lewis} and introduce the $\chi^2$ statistics as
\begin{equation}
\label{new9} \chi^2 = \min_{\gamma} \sum_{ij}
\left( \mu_i^{\rm exp} - \gamma - \mu_i \right) \sigma^{-2}_{ij} \left( \mu_j^{\rm exp} - \gamma - \mu_j \right)\ ,
\end{equation}
where the double sum runs over the 557 SNe, $\mu_i^{\rm exp}$ is the experimental value of the distance modulus
of the $i^{\rm th}$ supernova, $\mu_i$ the corresponding theoretical value given by Eq.~(\ref{mu}),
$\sigma^{2}_{ij}$ is the covariance matrix (containing both statistical and systematic errors),
and $\gamma$ an unknown normalization parameter~\cite{Melchiorri}.
The values of $\mu_i^{\rm exp}$ and $\sigma^{2}_{ij}$ in the our analysis are taken from Ref.~\cite{Union2}.

%************************************   Figure 6  *******************************************%
%
\begin{figure*}
\begin{center}
\includegraphics[clip,width=20cm,bb=50 40 570 350]{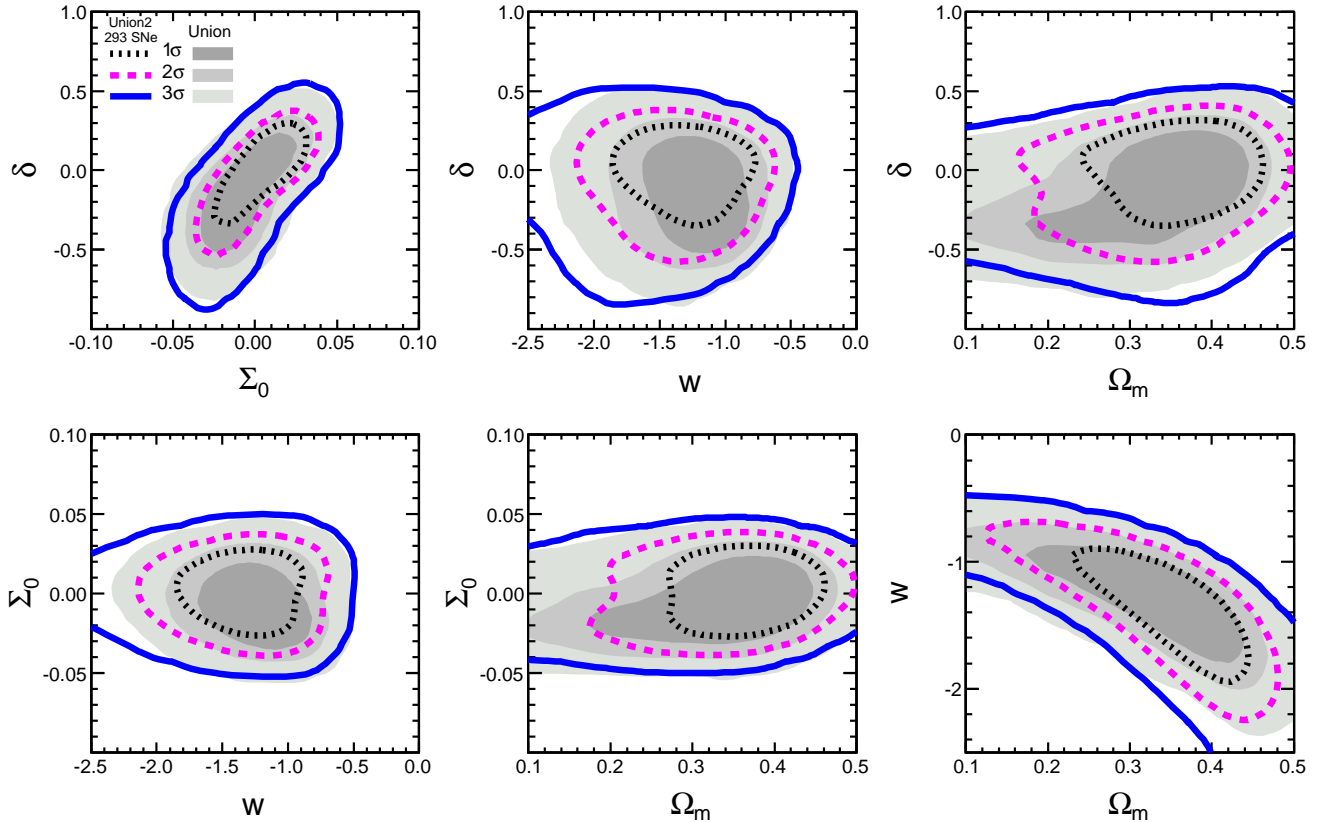}
\caption{$1\sigma$, $2\sigma$, and $3\sigma$ confidence
level contours in the planes $(\Sigma_0,\delta)$, $(w,\delta)$, $(\Omega_m,\delta)$,
$(w,\Sigma_0)$, $(\Omega_m,\Sigma_0)$, and $(\Omega_m,w)$.
Gray filled contours come from the Union data analysis detailed in Section IV.a,
while empty contours refer to the Union2 data analysis %described in Section IV.b
of just the 293 SNe contained in both compilations.}
\end{center}
\end{figure*}
%
%*********************************************************************************************%

The minimum in Eq.~(\ref{new9}) can be evaluated analytically and gives
\begin{equation}
\label{new10} \chi^2 = \sum_{ij}
\left( \mu_i^{\rm exp} - \tilde{\mu}_i \right) \left( \sigma^{-2}_{ij} - M_{ij} \right)
\left( \mu_j^{\rm exp} - \tilde{\mu}_j \right)\ ,
\end{equation}
where we have introduced the matrix
\begin{equation}
\label{new11} M_{ij} = \frac{\sum_{kl} \sigma^{-2}_{ik} \sigma^{-2}_{lj}}{\sum_{kl} \sigma^{-2}_{kl}}
\end{equation}
and the ``reduced'' distance modulus
\begin{equation}
\label{new12} \tilde{\mu} = 5 \, \log_{10} \tilde{d}_L  + 25 \ ,
\end{equation}
with the ``reduced'' luminosity distance
\begin{equation}
\label{new13} \tilde{d}_L = H_0 d_L \ .
\end{equation}
It is worth noticing that $\tilde{d}_L$ is independent on
the Hubble parameter $H_0$, so both $\tilde{\mu}$ and the $\chi^2$ in Eq.~(\ref{new10}) depend only
on the cosmological parameters $(\Omega_m,\Sigma_0,w,\delta,b_A,l_A)$.
Since the $\chi^{2}$ depends on as many as six parameters, we employ a
Markov Chain Monte Carlo approach.

%*************************************   Table II   ******************************************%

\begin{table}[t]
\vspace*{-0.2cm}
\begin{center}
\caption{Best-fit (BF) values, and the $1\sigma$ and $2\sigma$ confidence level intervals
derived from the Union2 data analysis, for
the cosmologically relevant parameters $\Sigma_0$, $\delta$, $w$, and $\Omega_m$.}
\vspace{0.3cm}
\begin{tabular}{llllllll}
\hline \hline
& &~~~~~~~\:$\Sigma_0$ &~~~~~~\;\;$\delta$ &~~~~~~~~$w$       &~~~~\,$\Omega_m$  \\
\hline
&BF &~~~\:$-0.004$       &~~~$-0.05$        &~~~~$-1.32$       &~~~~$0.37$        \\
&$1\sigma$ &$[-0.012,0.012]$    &$[-0.16,0.12]$     &$[-1.66,-1.04]$    &$[0.29,0.41]$     \\
&$2\sigma$ &$[-0.024,0.046]$    &$[-0.29,0.246]$     &$[-1.96,-0.79]$    &$[0.22,0.46]$     \\
\hline \hline
\end{tabular}
\end{center}
\vspace*{-0.1cm}
\end{table}

%*********************************************************************************************%

{\it Results.}
Figure~3 (empty contours) shows the $1\sigma$, $2\sigma$, and $3\sigma$ contours in the planes
$(\Sigma_0,\delta)$, $(w,\delta)$, $(\Omega_m,\delta)$, $(w,\Sigma_0)$, $(\Omega_m,\Sigma_0)$,
and $(\Omega_m,w)$.
Figure~5, instead, shows the $\sqrt{\Delta \chi^2}$ distributions for $\Sigma_0$, $\delta$, $w$,
and $\Omega_m$, the two distributions for $l_A$ and $b_A$ being flat (not shown).
Finally, Table~II shows the best-fit values and the $1$ and $2\sigma$ confidence level intervals
for the same parameters.
The $\chi^2$ at the best-fit point is 528.9.

The Union2 data analysis gives more stringent bounds on the anisotropy parameters
with respect to the Union analysis. In fact, from the top-left panel of Fig~3,
it can be seen that the uncertainties on $\Sigma_0$ and $\delta$ are reduced
by about one half.
Moreover, going from Union to Union2 analysis, the uncertainty on $w$ is practically unchanged,
while slightly higher values of $\Omega_m$ are preferred.

However, both the anti-correlation between $w$ and $\Omega_m$ and the correlation
between $\delta$ and $\Sigma_0$ still persist, as well as the weak
dependence of $\delta$ and $\Sigma_0$ on $w$ and $\Omega_m$.

%************************************   Figure 7  *******************************************%
%
\begin{figure*}
\begin{center}
\hspace*{-0.57cm}
\includegraphics[clip,width=8.65cm,bb=100 40 490 410]{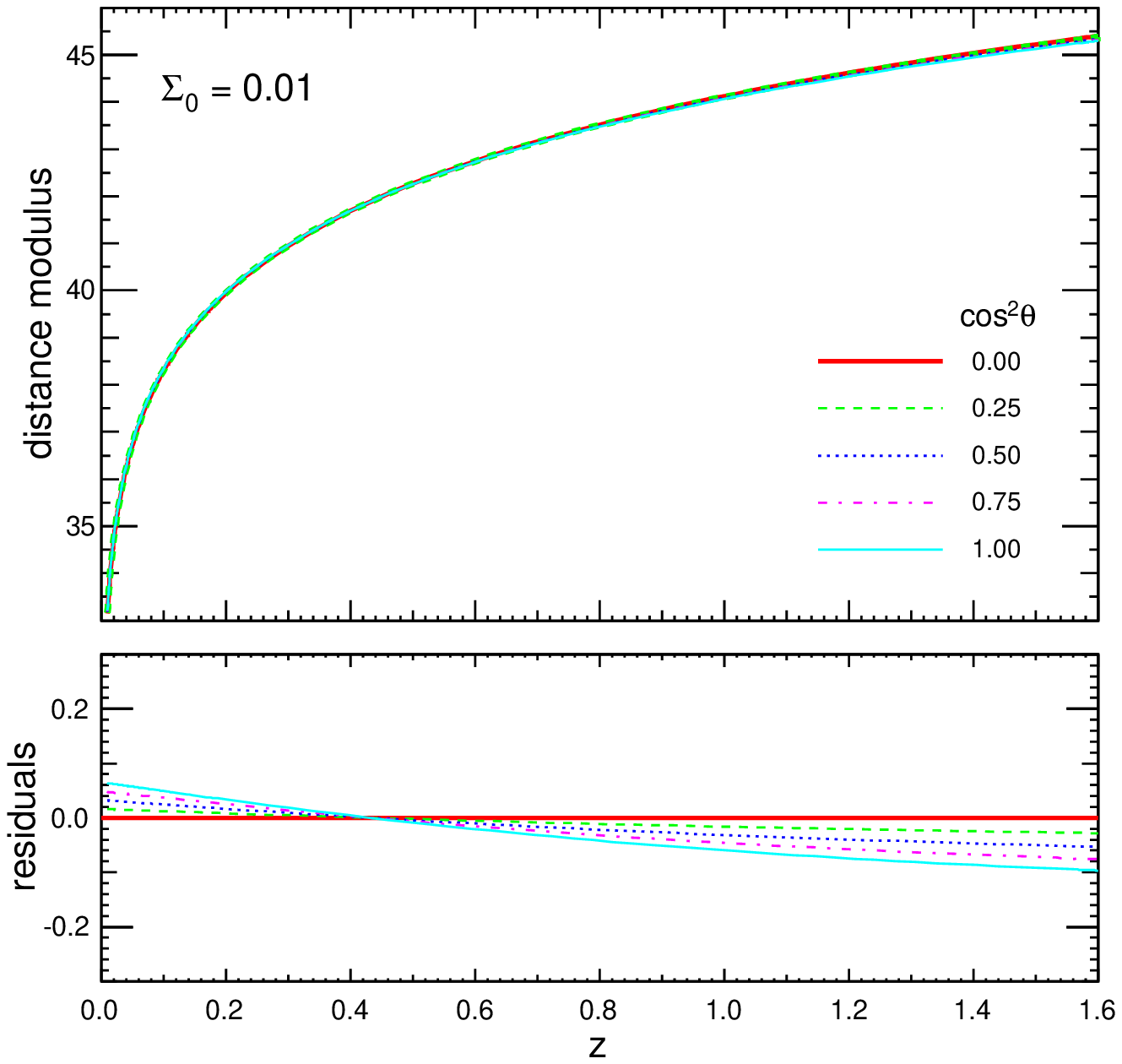}
\hspace*{0.15cm}
\includegraphics[clip,width=8.65cm,bb=100 40 490 410]{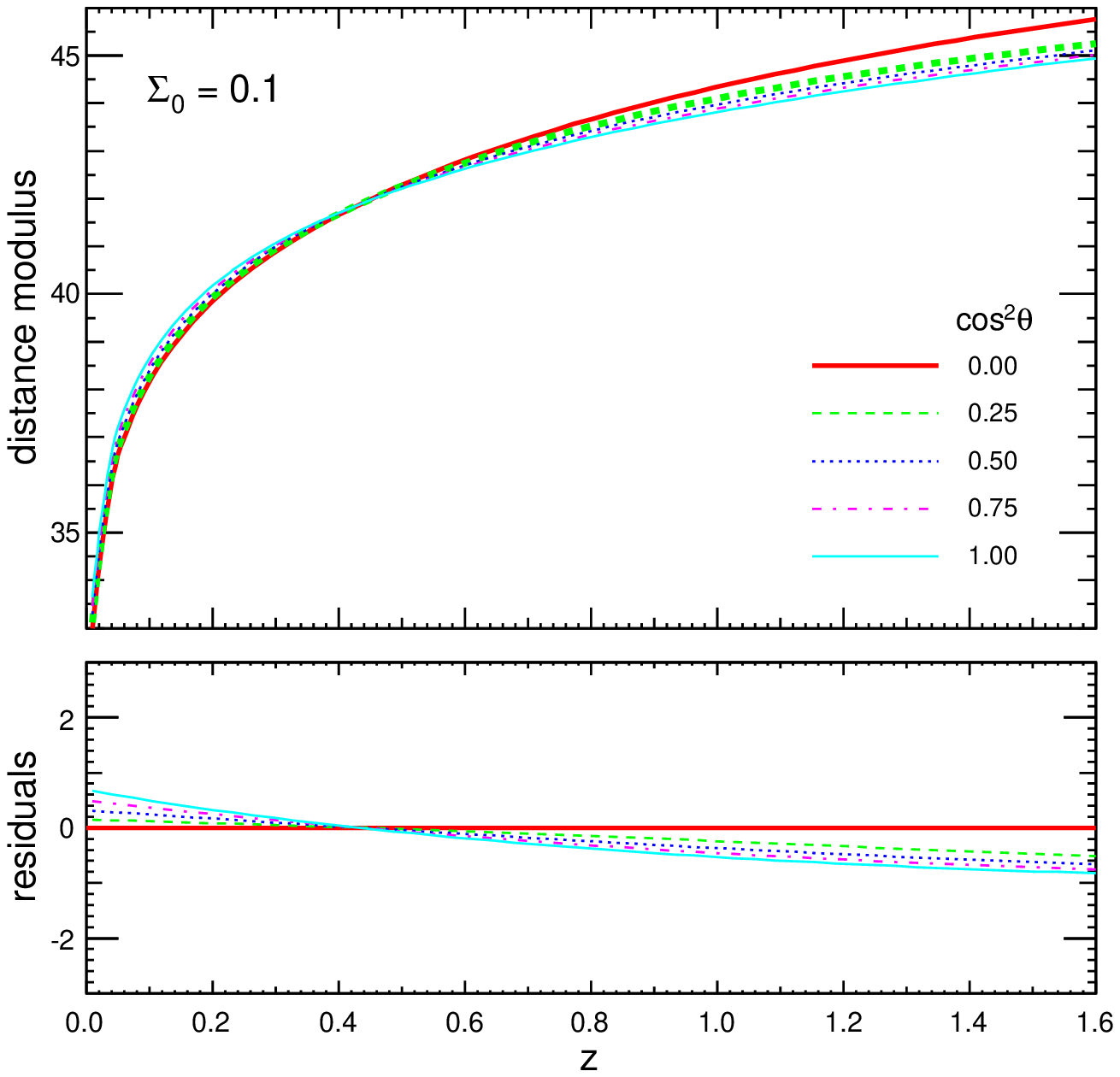}
\caption{{\it Left panels.} Upper panel: Distance modulus as a function of the redshift for different values
of $\cos^2 \! \theta$, with $\theta$ being the angle between the anisotropy axis and
the generic direction of a SN, for $\Sigma_0 = 0.01$.
The remaining parameters are fixed to
$(H_0,\delta,\Omega_m,w) = (71\,\mbox{km}/\mbox{s}/\mbox{Mpc},10\Sigma_0,0.3,-1)$.
Lower panel: Residuals (distance modulus minus distance modulus for the case
$\cos^2 \! \theta = 0$) for the same models in the upper panel.
{\it Right panels.}  The same as the left panels but with $\Sigma_0 = 0.1$.}
\end{center}
\end{figure*}
%
%*********************************************************************************************%

In order to check if the reduction of $\Sigma_0$ and $\delta$ uncertainties
is due to the increase of statistics in the Union2 compilation with respect to Union
(557 compared to 307 SNe) or to the the different method of analysis of SN data,
we repeated the Union2(-like) analysis keeping only those 293 SNe which appear in both
compilations. The results are shown in Fig.~6, superimposed to the previous Union fit.
The $\chi^2$ at the best-fit point is 263.2.

A part from slightly higher values of $\Omega_m$ and $\delta$ preferred by the 293 SN
Union2 fit, the two different analyses are in very good agreement. This fact highlights
the robustness of our bounds on the cosmic anisotropy obtained with type Ia supernovae.

As a final check of correctness of our analysis in determining the amount of anisotropies,
we verified that our results do not change if we exclude from the analysis supernovae at low
redshifts, which could eventually give spurious values of anisotropies on far larger scales.
In particular, we cut first SNe with $z \leq 0.02$ and then SNe with $z \leq 0.03$,
finding no appreciable changes in the confidence level intervals for
$\Sigma_0$, $\delta$, $w$, and $\Omega_m$ (not shown).

\subsection{\normalsize{IV.c Anisotropy Axis}}
\renewcommand{\thesection}{\arabic{section}}

We have found, in the previous Sections, no evidence for a preferred axis of symmetry,
since the probability distributions for $l_A$ and $b_A$ are flat.

This is due to the fact
that the allowed values of the cosmic shear $\Sigma_0$ are relatively small,
$|\Sigma_0| \lesssim 0.01$ at $1\sigma$ C.L. (see Table~II).

Indeed, for small values of the shear, the distance modulus Eq.~(\ref{mu}) weakly depends
on both the angular position of SNe, $\hat{n}$, and on the direction of the symmetry axis,
$\hat{n}_A$.

More precisely, these two angular positions enter in the expression of
$\mu$ just through the particular combination
$\cos^2 \theta = (\hat{n} \cdot \hat{n}_A)^2$ [see Eqs.~(\ref{eccentricityA})-(\ref{mu})], and the
distance modulus is not much sensible to the variations of $\cos^2 \theta$
when $|\Sigma_0|$ is small.

This can be appreciated in Fig.~7, where we show $\mu$ as
a function of the redshift for different values
of $\cos^2 \! \theta$ in the two cases $\Sigma_0 = 0.01$ (left panel) and
$\Sigma_0 = 0.1$ (right panel), with $\delta = 10\Sigma_0$ and
assuming the fiducial cosmology $(H_0,\Omega_m,w) = (71\,\mbox{km}/\mbox{s}/\mbox{Mpc},0.3,-1)$.

The variation of $\mu$ for $\Sigma_0$ as small as $0.01$ is well within the experimental
uncertainty on the distance modulus (of about 0.3 magnitudes), so that a determination of the symmetry axis is
beyond the actual experimental possibility. On the other hand, for values of the shear
as large as $0.1$ a determination of $\hat{n}_A$ is in principle possible --due to the significant
dependence of $\mu$ on $\cos^2 \! \theta$-- but such large values of $\Sigma_0$ are
excluded by data.

Therefore, we conclude that SN data are not able to constrain, at the same time, the
anisotropy parameters ($\Sigma_0$ and $\delta$) and the preferred direction defined
by the anisotropy itself.

However, it is worth noticing that the approach detailed in this paper (that provides bounds
on $\Sigma_0$ and $\delta$ but not on $\hat{n}_A$) can be
confronted with the so-called ``hemisphere comparison method''~\cite{Schwarz} (which searches
for a preferred axis in magnitude-redshift data without constraining the
level of anisotropy) to give a better a complete understanding of the geometry of the Universe
on large scales.

Indeed, the hemisphere comparison method applied to Union2 data favors
an anisotropic accelerated expansion in the direction~\cite{Antoniou}
$(b_A,l_A) = \left(18^{\circ \, +11^\circ}_{\;\;\,-10^\circ},309^{\circ \, +23^\circ}_{\;\;\,-3^\circ}\right)$,
which is consistent, both qualitatively and quantitatively, with the evidence of a
preferred direction in the sky coming from other cosmological observations (namely,
large-scale velocity flow axis, alignment of CMB quadrupole and octupole axes,
and quasar optical polarization alignment axis, all discussed in Ref.~\cite{Antoniou}).

%***********************************   Conclusions   *****************************************%

\section{\normalsize{V. Conclusions}}
\renewcommand{\thesection}{\arabic{section}}

The analysis of cosmic microwave background radiation reveals that the level of anisotropy
in the large-scale geometry of the Universe must be below
$10^{-5}$~\cite{parallax} at redshifts $z \sim 10^3$.
This constraint do not exclude that a much larger level of anisotropy is allowed at recent times,
$z \sim \mathcal{O}(1)$, as first noted in Ref.~\cite{Jimenez} and Ref.~\cite{Koivisto-Mota}.

In particular the authors of Ref.~\cite{Koivisto-Mota}
analyzed the magnitude-redshift data on type Ia supernovae in the ``GOLD'' data set of Riess et
al.~\cite{Riess2006} (consisting of 182 supernovae), in a nonstandard cosmological model with anisotropic
dark energy. They found that deviations from isotropy in the equation of state of dark energy
were constrained at the level of $|\delta| < few \times 10^{-1}$.

In this paper, we have extended the work~\cite{Koivisto-Mota} by constraining also the present level of
cosmic anisotropy $\Sigma_0$, namely, the anisotropy in the large-scale geometry of the Universe, and
by analyzing the magnitude-redshift data on type Ia supernovae in the Union and Union2 data sets of
Kowalski et al.~\cite{Kowalski} (consisting of 307 supernovae) and
Amanullah et al.~\cite{Union2} (consisting of 557 supernovae), respectively.
In particular, by using Union2 data, we have confirmed the results of~\cite{Koivisto-Mota}
about the skewness, finding
\begin{equation}
-0.16 < \delta < 0.12 \;\; (1\sigma \; \mbox{C.L.})\ ,
\end{equation}
and we have put first limits on the present cosmic shear parameter $\Sigma_0$,
\begin{equation}
-0.012 < \Sigma_0 < 0.012 \;\; (1\sigma \; \mbox{C.L.})\ .
\end{equation}
We conclude that a standard isotropic universe is consistent with SN data,
any deviation at redshifts $z \lesssim 1.6$ being constrained by the above results.

%*********************************   Acknowledgments   ***************************************%

\begin{acknowledgments}
We would like to thank E. Lisi for useful discussions and valuable comments.
\end{acknowledgments}

%**********************************   Bibliography   *****************************************%

\end{document}